\documentclass[11pt,showpacs,aps]{revtex4}

\usepackage{graphicx}

\begin{document}

\preprint{}

\title{Gravitational instanton and cosmological term}

\author{She-Sheng Xue}

\email{xue@icra.it}

\affiliation{ICRANet and
Physics Department, University of Rome ``La Sapienza", 00185 Rome, Italy}



\begin{abstract}
Quantum fluctuation of unstable modes about gravitational instantons causes 
the instability of flat space at finite temperature, leading to the 
spontaneous process of nucleating quantum black holes. The energy-density of quantum black holes, 
depending on the initial temperature, 
gives the cosmological term, which naturally accounts for the inflationary phase of Early Universe. 
The reheating phase is attributed to the Hawking radiation and annihilation of these quantum 
black holes. Then, the radiation energy-density dominates over the energy-density of quantum black holes,
the Universe started the Standard cosmology phase. In this phase the energy-density of quantum black holes
depends on the reheating temperature. It asymptotically approaches to the cosmological constant 
in matter domination phase, consistently with current observations.    
\end{abstract}


\maketitle

\section{ Introduction.} 

In recent years, as observational data concerning on cosmology are rapidly accurate, 
the theoretical understanding of our Universe has been greatly profounded ever before. 
The inflation \cite{linde} in early Universe and the acceleration of present 
Universe \cite{wein} are two most important issues in modern cosmology and fundamental 
physics. Both issues are closely related to the cosmological term in Einstein equation.
The observational value of the cosmological constant  
\begin{equation}
\Lambda^{\rm obs} = 8\pi G \epsilon_\Lambda \sim 10^{-118}m_p^2,
\label{lambdaobs}
\end{equation}
where the Newton constant $G=1/m_p^2$, the Planck mass $m_p\sim 10^{19}$GeV and a related characteristic 
energy-density
\begin{equation}
\epsilon_\Lambda \sim 10^{-120}m_p^4.
\label{edobs}      
\end{equation}
This energy-density is of $120$ order smaller than the Planck energy-density,
$\epsilon_{\rm Planck} = m_p^4$. 

Much effort in understanding these issues has been made for many decades 
and there are many interesting ideas and innovative theoretical developments based 
on either simple models \cite{q} or complex theories \cite{s}. In these approaches, 
a scalar field slowly-rolling downwards an effective potential, 
which mimics a possible vacuum-energy variation, plays a crucial r\^ole in driving 
inflation; whereas this scalar field with a mass of order the current Hubble scale, can 
possibly account for acceleration. 

In this article, we attempt to study
these two issues within a framework based on the instability of flat space at finite temperature,
due to an unstable quantum fluctuation about gravitational instantons.
We shall use the signature $(- + + +)$.

\section{Cosmological term and energy-momentum tensors}

\subsection{Cosmological constant term}

The gravitational field interacts only with the total energy-momentum ${\mathcal T}_{ab}$ that is attributed to 
both fundamental particles and vacuum fluctuations.  
The total energy-momentum tensor ${\mathcal T}_{ab}$ appears as source of 
the gravitational field. The gravitational field equation should therefore be of form 
\begin{equation}
{\mathcal G}_{ab}= 8\pi G {\mathcal T}_{ab},
\label{geq0}
\end{equation}  
where $G$ is the Newton constant. In the left-handed side of the field equation, 
${\mathcal G}_{ab}$ is some expression involving the metric $g_{ab}$ 
and its derivatives. The field equation (\ref{geq0}) is invariant under general coordinate transformations, 
therefore ${\mathcal G}_{ab}$ must be a tensor. It is further assumed that the field equation (\ref{geq0}) is of 
second differential order, and is linear in the second derivatives. We 
know that the only tensor that contains the second derivative linearly is $R_{abcd}$. 
Hence the only tensors available for the construction of ${\mathcal G}_{ab}$ are $R_{abcd}$ and $g_{ab}$. 
The most general second-rank tensor that can be built out of these, with $R_{abcd}$ entering linearly, is    
\begin{equation}
{\mathcal G}_{ab}=  R_{ab} +B g_{ab}R +\Lambda g_{ab},
\label{gab}
\end{equation}
where $R_{ab}$ ($R$) is the Ricci tensor (scalar), $B$ and $\Lambda$ are constants.  
The conservation law for the 
energy-momentum tensor of matter fields ${\mathcal T}^b_{a;b}=0$ requires the object ${\mathcal G}_{ab}$ must 
satisfy the identity
\begin{equation}
{\mathcal G}^b_{a;b}=0.
\label{geqi0}
\end{equation}
In view of $g^b_{a;b}=0$ and the Bianchi identity
\begin{equation}
( R^b_a-\frac{1}{2}\delta^b_a)_{;b}=0,
\label{bianchi}
\end{equation}
the identity (\ref{geqi0}) is fulfilled if $B=1/2$. The $\Lambda$ remains as an arbitrary constant. 
Thus, gravitational field equation becomes
\begin{equation}
R_{ab}-\frac{1}{2}g_{ab}R + \Lambda g_{ab}=8\pi G {\mathcal T}_{ab},
\label{e1}
\end{equation} 
where $\Lambda g_{ab}$ is the cosmological term ($\Lambda$-term) and $\Lambda$ is the so-called 
cosmological constant. The $\Lambda$-term
clearly has its gravitational origin (or geometric origin), rather than material origin. In Newtonian gravitating systems, the cosmological constant $\Lambda=0$
consistently with observations and experiments. 

\subsection{Energy-momentum tensors of fundamental particles and vacuum}\label{vacuum0}

In the context of local quantum field theories, the total energy-momentum tensor ${\mathcal T}_{ab}$, 
the right-handed side of the gravitational field equation (\ref{geq0},\ref{e1}), can be written as,
\begin{equation}
{\mathcal T}_{ab} = \langle 0| \left(\hat T_{ab} +\hat T^V_{ab}\right)|0\rangle 
+ \langle 0| \Delta\hat T^V_{ab} |0\rangle,
\label{gt}
\end{equation}
where $|0\rangle$ stands for the vacuum state and $\langle 0|\cdot\cdot\cdot |0\rangle$ denotes 
the vacuum expectational value. $\hat T_{ab}$ represents the energy-momentum of fundamental 
particles (including gravitational field) and 
$\hat T^V_{ab}$ represents the energy-momentum of the vacuum.
$ \Delta\hat T^V_{ab}$ represents the energy-momentum contribution from particles back 
interacting on the vacuum. 

Only energy-momentum difference between quantum states can be physically significant in the flat space time. 
As a consequence, the vacuum is 
defined as the ground state with zero energy-momentum, 
\begin{equation} 
\langle 0|\hat T^V_{ab}|0\rangle \equiv 0
\label{cgt}
\end{equation}
and fundamental particles are energy-momentum excitations 
from the vacuum. The energy-momentum of fundamental particles is then given by
\begin{equation} 
T_{ab}=\langle 0| \left(\hat T_{ab} +\hat T^V_{ab}\right)|0\rangle =\langle 0|\hat T_{ab}|0\rangle.
\label{cgt1}
\end{equation}
The $T_{ab}$ is in fact the {\it net} contribution from fundamental particles and the vacuum in 
the context of quantum field theories. It corresponds the classical energy-momentum,
entering the right-handed side of the field equation (\ref{e1},\ref{gt}) as the source for gravitational field. 
 
However, we expect to have some problems in such a scenario: 
(i) the ground state (vacuum) of quantum gravity at the Planck scale is still an open question; 
(ii) the energy-momentum difference (\ref{cgt}) between fundamental particles 
states and vacuum state must be altered by
the back reaction of fundamental particles (including gravitational field) on the vacuum state; 
(iii) at very high temperature $T \sim m_p$,
thermal fluctuations are much more important than quantum-field fluctuations, and the vacuum is not stable.    
       
\subsection{Back reactions on the vacuum}

The second part $\langle 0|\Delta \hat T^V_{ab} |0\rangle$ in the total energy-momentum tensor 
${\mathcal T}_{ab}$ (\ref{gt}) is the variation of the vacuum energy-momentum, 
\begin{equation} 
\langle 0|\Delta \hat T^V_{ab} |0\rangle = \langle 0|\hat T^V_{ab}|0\rangle_B -\langle 0|\hat T^V_{ab}|0\rangle,
\label{deltat}
\end{equation}
where $\langle 0|\hat T^V_{ab}|0\rangle_B$ is the modified energy-momentum tensor caused by back reactions
of external sources: (i) boundary conditions; (ii) classical fields; and (iii) fundamental particles 
on the vacuum.  

The vacuum energy-momentum is altered, due to external sources back interacting 
with vacuum fluctuations. The vacuum fluctuations, whose wavelengths are the order of the characteristic 
length scale of external sources, are most strongly affected. As a result, the variation of 
vacuum energy-momentum $\langle 0|\Delta T^V_{ab} |0\rangle$ is determined by the characteristic scales 
of external sources. In the case of external source being boundary condition, for example 
the Casimir effect \cite{casimir}, the variation 
of vacuum energy-density is  
\begin{equation} 
\Delta \epsilon^{\rm Casimir}_0 = -\frac{\pi^2}{720}\frac{1}{l_{\rm C}^4},
\label{casimirv}
\end{equation}
where the characteristic scale is $l_{\rm C}$ is the separation between two conducting plates. 
In the case of external source being 
classical gravitational field, for example the Hawking effect \cite{hawking,xue2} 
for an astrophysical black hole, the variation of vacuum energy-density is (a scalar field with two polarizations)  
\begin{equation} 
\Delta \epsilon^{\rm Hawking}_0 = -\frac{\pi^2}{45}T_H^4,
\label{hawkingv}
\end{equation}
where the Hawking temperature $T_H=1/(8\pi GM)$, $M$ is the black hole mass and the characteristic scale is the 
black hole size $GM$. With macroscopic lengths 
$l_{\rm C}> 10^{-4}$cm and 
astrophysical black hole sizes $GM > 10^5$cm, one finds that the variations of vacuum energy-density are very much
smaller than the energy-density $m_e^4$ of an electron at rest, where $m_e$ is the electron mass. 
The variation of the vacuum energy-momentum $\langle 0|\Delta \hat T^V_{ab} |0\rangle$ 
due to back reactions has very small gravitating effect and is negligible in the right-handed 
side of gravitational field equation (\ref{e1}).

In the assumption that vacuum fluctuations are homogeneously distributed in
whole Universe, it is conceivable that the variations of vacuum energy-momentum induced by back reactions of 
external sources: topology boundaries of the Universe and classical gravitational perturbations 
at the cosmological scale ($\sim 10^{28}$cm) are extremely small,
\begin{equation} 
\Delta \epsilon^{\rm Cosmo}_0 \sim 10^{-164}[{\rm GeV}]^4.
\label{cosmov}
\end{equation} 
Comparing it with the energy-density $\epsilon_\Lambda$ (\ref{edobs}), one finds that Eq. (\ref{cosmov})
is of $117$ order smaller.  Therefore the variations of vacuum energy-momentum due to back reaction 
have entirely negligible impacts on the evolution of the Universe governed by gravitational field 
equation (\ref{e1}). 

In conclusion, the vacuum energy-momentum (\ref{cgt}) $\langle 0|\hat T^V_{ab} |0\rangle\equiv 0$ 
and the variation of the vacuum energy-momentum (\ref{deltat}) 
$\langle 0|\Delta \hat T^V_{ab} |0\rangle\simeq 0$, the total energy-momentum tensor (\ref{gt}) becomes
\begin{equation}
{\mathcal T}_{ab}\simeq T_{ab}. 
\label{gtf}
\end{equation}
The gravitational field equation (\ref{e1}) can be written as 
the Einstein equation,
\begin{equation}
R_{ab}-\frac{1}{2}g_{ab}R + \Lambda g_{ab}=8\pi G T_{ab}.
\label{e2}
\end{equation}
The cosmological term is clearly distinguished from the energy-momentum tensor of fundamental particles, 
and they play different r\^oles in either side of the Einstein equation (\ref{e2}). It can be 
conceptually misleading, if one simply moves the $\Lambda$-term from the left-handed side to the right-handed
side of the Einstein equation (\ref{e1}). 

\subsection{Cosmological term $\not=$ vacuum-energy term}

One is certainly allowed to optionally adopt a non-vanishing vacuum energy-momentum tensor 
$\langle 0|\hat T^V_{ab}|0\rangle\not=0$,
and model the vacuum of quantum field theories 
as a perfect fluid with energy-density $\epsilon_0$, pressure $p_0$ and the equation of state $\epsilon_0 = -p_0$, 
\begin{equation} 
\langle 0|\hat T^V_{ab}|0\rangle = p_0g_{ab}-(\epsilon_0 + p_0 )u_a u_b = -\epsilon_0 g_{ab},
\label{vfluid}
\end{equation}
where $u_a$ is the four velocity of fluid elements. However, in this framework, the 
energy-momentum tensor $\langle 0|\hat T_{ab}|0\rangle$ in Eq. (\ref{cgt1}) is no longer the classical
energy-momentum tensor $T_{ab}$ of fundamental particles,
\begin{equation} 
\langle 0|\hat T_{ab}|0\rangle =\tilde T_{ab}\not= T_{ab},
\label{vtn1}
\end{equation}
and the total energy-momentum (\ref{gt})is 
\begin{equation} 
\quad {\mathcal T}_{ab} \simeq \langle 0| \left(\hat T_{ab} +\hat T^V_{ab}\right)|0\rangle 
=\tilde T_{ab}-\epsilon_0 g_{ab},
\label{vtn2}
\end{equation}
where the back reaction term $\langle 0|\Delta \hat T^V_{ab}|0\rangle$ is neglected. The Einstein equation becomes
\begin{equation}
R_{ab}-\frac{1}{2}g_{ab}R + \Lambda g_{ab}=8\pi G \left(\tilde T_{ab}-\epsilon_0 g_{ab}\right).
\label{e3}
\end{equation}
It is completely not justified that one simply 
lets $\tilde T_{ab}=T_{ab}$, moves the cosmological term $\Lambda g_{ab}$ to the right-handed side 
of the Einstein equation and relates it to the vacuum energy-density,
\begin{equation}
R_{ab}-\frac{1}{2}g_{ab}R =8\pi G \left(T_{ab}-\frac{\Lambda}{8\pi G} g_{ab}\right),\quad \Lambda =8\pi G \epsilon_0.
\label{e4}
\end{equation}  
The problem of conceptually confusing the cosmological term and vacuum energy in the Einstein equation
is much worst than the problem of large order magnitude discrepancy between the 
observed energy-density (\ref{edobs}) and the Planck energy-density $m_p^4$.
        
In the empty space, there are no any fundamental particles $T_{ab}=0$, and 
the cosmological term is set to be zero, the vacuum Einstein equation (\ref{e2}) reads 
\begin{equation}
R_{ab}-\frac{1}{2}g_{ab}R =0.
\label{e5}
\end{equation}
We attempt to study a possible origin of the cosmological term from quantum (thermal) 
fluctuations about gravitational instanton by using semi-classical saddle-point 
approach to the functional integral for quantum gravity.

\section{Gravitational instantons and functional integrals}

The attractive nature of gravity that cannot be screened is the essential reason for
many inevitable instabilities of classical gravitating systems, such as the gravitational collapse and 
Jean instability. One might also worry about the instability of classical solutions of the Einstein 
equation against quantum fluctuations. 
To explore the structure of quantum (thermal) fluctuations about these classical solutions of the vacuum
Einstein equation (\ref{e5}), one starts with the Euclidean 
action $S_E(g)$ and functional integral for quantum gravity \cite{gross},
\begin{equation}
Z=\int {\mathcal D}[g_{ab}(x)]e^{-S_E(g)+ {\rm gauge-fixing}\hskip0.05cm {\rm terms}},
\label{function0}
\end{equation}
where the Euclidean action is given by,
\begin{equation}
S_E(g)=-\frac{1}{16\pi G}\int (g)^{1/2}d\tau d^3x R + {\rm boundary}\hskip0.15cm {\rm terms},
\label{action0}
\end{equation}
where we make the Wick rotation, the Euclidean time $d \tau = -id t$ and 
signature $(+ + + +)$, $(g)^{1/2}d\tau d^3x$ is the Euclidean volume element. 
Boundary terms are added to have an action which reproduces 
the vacuum Einstein equations under all variations of the metric that vanishes on the boundary. 
The functional integration is evaluated by integrating over all metrics $g_{ab}$ that are positive 
and definite and obey appropriate boundary conditions. The functional integral (\ref{function0})
is treated by saddle-point methods. This is adequate for 
semi classical analysis of small quantum (thermal) fluctuations about classical solutions of the vacuum
Einstein equation, in order to study the stability of these solutions against small quantum (thermal) fluctuations.  

The saddle-point evaluation of the functional integral $Z$ (\ref{function0}) starts 
by constructing the saddle-point of the action, namely, the classical solutions $g_{ab}^s$ 
of the vacuum Einstein equation (\ref{e5}) and $R=0$. Expansion about these saddle points is performed 
by writing
\begin{equation}
g_{ab}(x)=g_{ab}^s(x)+\phi_{ab}(x),
\label{fluctuations}
\end{equation}
where $\phi_{ab}$ are small quantum (thermal) fluctuation fields about $g_{ab}^s$ as $c$-number background fields. 
The saddle-point metric $g_{ab}^s$ (normally assumed to be nonsingular geodesically complete four-manifold)
is colloquially termed a gravitational instanton \cite{gross}.
Two types of boundary conditions that gravitational instantons $g^s_{ab}(x)$ obey are discussed 
in Ref. \cite{gross}. One corresponds to the zero-temperature vacuum and another corresponds 
to the canonical ensemble at temperature $T=1/\beta$. 

The boundary conditions appropriate to the 
zero-temperature vacuum are termed by asymptotically Euclidean (AE) \cite{gross27}. An AE metric is one in which 
the metric approaches the flat metric on $R^4$ outside some compact set. Finite action requires the metric 
to be asymptotically,
\begin{equation}
ds^2=\left(1+\frac{\alpha}{r}\right)\delta_{ab}dx^adx^b+O(r^{-3}),
\label{ae}
\end{equation}
where $r$ is a four-dimensional radial coordinate and $\alpha$ a function of coordinates, but independent of 
$r$. The boundary at infinity is topologically $S^3$. The gravitational instantons $g^s_{ab}(x)$ obeyed such 
boundary condition (\ref{ae}) are colloquially termed gravitational AE instantons. 

The boundary conditions for the canonical ensemble at temperature $T=1/\beta$ are termed by asymptotically flat (AF) \cite{gross27}. An AF metric is one in which the metric approaches the flat metric on $R^3\times S^1$ outside
some compact set. Finite action requires the metric to be asymptotically,
\begin{equation}
ds^2=d\tau^2+\left(1+\frac{\alpha}{|{\bf r}|}\right)\delta_{ij}dx^idx^j+O(|{\bf r}|^{-3});\quad i,j=1,2,3,
\label{af}
\end{equation}
where $|{\bf r}|$ is a three-dimensional radial coordinate, $\alpha$ a function of coordinates, but
independent of $|{\bf r}|$; the Euclidean time $\tau$ is a coordinate which is periodic with period $\beta$. The
boundary of infinity is topologically $S^2\times S^1$.
The gravitational instantons $g^s_{ab}(x)$ obeyed such 
boundary condition (\ref{af}) are colloquially termed gravitational AF instantons.

In Eqs. (\ref{function0},\ref{fluctuations}), treating $\phi_{ab}$ as small quantum (thermal) fluctuation fields and 
instantons $g_{ab}^s$ as $c$-number background fields will generate the usual perturbation expansion. 
Up to the lowest order, i.e., the quadratic terms of quantum fields $\phi_{ab}$, one finds
\begin{eqnarray}
Z(g^s) &\approx & e^{-S_E(g^s)}\int {\mathcal D}\phi^\dagger{\mathcal D}\phi
\exp\left[-\frac {1}{2}\int (g)^{1/2}d\tau d^3x \phi^\dagger {\mathcal M}(g^s)\phi\right]\nonumber\\
&=&\exp\left\{-S_E(g^s)-\frac{1}{2}\ln\det[{\mathcal M}(g^s)]\right\},
\label{functional}
\end{eqnarray}
where metric indexes are omitted. $S_E(g^s)$ is the action of gravitational AE or AF instantons. 
The functional integration is evaluated by integrating over the quantum (thermal) fluctuation fields $\phi_{ab}$
about the saddle-points $g_{ab}^s$. The matrix ${\mathcal M}(g^s)$ determines the properties of quantum 
fluctuation fields $\phi_{ab}$ about instanton $g_{ab}^s$.

\section{Quantum (thermal) fluctuations about flat space}\label{quantumsec}

\subsection{The flat space at zero-temperature}\label{zerosec}

In the zero-temperature vacuum case, the positive-action theorem due to Schoen and Yau \cite{yao} 
states that for any AE metric $g^s$ with $R=0$, the action $S_E(g^s)$ is non-negative and $S_E(g^s)=0$ if
only if $g_{ab}^s$ is flat. It was shown \cite{gross} that the action for any AE instanton must be zero, 
which follows from the fact that any AE instanton will be a solution of $R_{ab}=0$. 
The positive-action theorem then guarantees that AE instanton must be a flat metric. For zero-temperature,
AE instanton is unique flat metric and one needs only examine the quantum fluctuations about flat space.
The operator ${\mathcal M}(g^s)$ in Eq. (\ref{functional}) is semi-positive definite \cite{gross}. 
Therefore the flat space in the zero-temperature vacuum
case is stable quantum mechanically as well as classically. This
precludes the possibility of flat space at zero temperature decaying by any mechanism.

\subsection{The flat space at non-zero temperature}\label{flatnonzero}

In the canonical ensemble at temperature $T=1/\beta$, the metrics $g_{ab}$ in Eq. (\ref{function0}) are
strictly periodic in Euclidean time $\tau$ with period $\beta$: $g_{ab}(\tau,x)=g_{ab}(\tau+\beta,x)$,
and integration over Euclidean time becomes, 
\begin{eqnarray}
\int d\tau = N_\beta \int_0^\beta d\tau,\quad \int (g)^{1/2}d^4x=N_\beta \int_0^\beta d\tau\int (g)^{1/2}d^3x,
\label{einteval}
\end{eqnarray}
and the Euclidean action (\ref{action0})
\begin{eqnarray}
S_E(g)=-\frac{N_\beta}{16\pi G} \int_0^\beta d\tau\int (g)^{1/2}d^3x R + {\rm boundary}\hskip0.15cm {\rm terms},
\label{eactiont}
\end{eqnarray}
where $N_\beta$ to be the number of periodic intervals in the Euclidean time.
Defining the measure of functional integral as,
\begin{eqnarray}
\int {\mathcal D}[g_{ab}(x)]
=\prod^{N_\beta}\int {\mathcal D}[g_{ab}(x)]_\beta,
\label{eintevalm}
\end{eqnarray}  
one rewrites the functional integral (\ref{function0},\ref{functional}) as the canonical partition function 
\begin{eqnarray}
Z(g^s)&=&\int {\mathcal D}[g_{ab}(x)]e^{-S_E(g)+ {\rm gauge-fixing}}\nonumber\\
&\approx & \prod^{N_\beta} e^{-S_E(g^s)}\int [{\mathcal D}\phi^\dagger{\mathcal D}\phi]_\beta
\exp\left[-\frac {1}{2}\int^\beta_0 d\tau\int (g)^{1/2}d^3x \phi^\dagger {\mathcal M}_\beta(g^s)\phi\right]\nonumber\\
&=&\prod^{N_\beta}\exp\left\{-S_E(g^s)-\frac{1}{2}\ln\det[{\mathcal M}_\beta(g^s)]\right\}.
\label{functionalf}
\end{eqnarray}
The classical solution $g^s$, quantum (thermal) fluctuation field $\phi$ and quadratic metric 
${\mathcal M}_\beta(g^s)$ are strictly periodic in Euclidean time $\tau$ with period $\beta$. Therefore, 
the canonical partition function (\ref{functionalf}) is the same for different periodic sectors and 
one can normalizes $N_\beta=1$.

One defines the free energy ${\mathcal F}(g^s)$ and the effective Euclidean action 
$S^{\rm eff}_E(g^s)$, to describe the equilibrium state of such a system, 
\begin{eqnarray}
Z(g^s)\equiv e^{\beta {\mathcal F}(g^s)}= e^{-S^{\rm eff}_E(g^s)};\quad S^{\rm eff}_E(g^s)= S_E(g^s)+\frac{1}{2}\ln\det[{\mathcal M}_\beta(g^s)],
\label{effeuclidean}
\end{eqnarray}
where $S^{\rm eff}_E(g^s)\equiv -\beta {\mathcal F}(g^s)$.
We will see that the temperature $T$ acts as an external " field ", and the real part and imaginary part 
of the effective Euclidean action describe the transition of space-time configurations. 
 
There is one periodic solution, namely, flat space $g^s_{ab}(x)=\delta_{ab}$:
\begin{equation} 
d^2s = d\tau^2 + (dx^2+dy^2+dz^2).
\label{flatspace}
\end{equation}
This is a trivial AF instanton, since it has zero action $S_E(g^s)=0$. 
The semi-positive eigenvalues of the operator ${\mathcal M}(g^s)$ represent the modes of
quantum (thermal) fluctuations
about the classical solution $g^s_{ab}(x)=\delta_{ab}$ (\ref{flatspace}), which is a strictly local 
minimum of the action. The contribution of these modes to the free energy ${\mathcal F}$, 
to the lowest order in Eq. (\ref{functionalf}), will simply be the free energy of an ideal gas 
of relativistic massless particles with two helicities at temperature $T$,
\begin{equation}
\frac{{\mathcal F}_0}{V}=-\frac{\pi^2 }{45}T^4,
\label{free0}
\end{equation}
the corresponding energy-density
\begin{equation}
\rho_0=\frac{\pi^2}{45}T^4,
\label{energy0}
\end{equation}
and the partition function
\begin{equation}
Z^0=\exp\left[\frac{\pi^2 }{45}T^3V\right],
\label{zfree0}
\end{equation}
i.e., the contribution from thermal gravitons in a box of volume $V$ at temperature $T$.
The higher-order [in $(16\pi G)^{1/2}$] corrections \cite{gross} to the free energy, arising from 
the self-interactions for the gas of gravitons, leads to the infrared instability, i.e., Jeans instability.

\section{Finite instanton action at non-zero temperature}\label{finitesec}

\subsection{Finite instanton action}

However, unlike the AE instanton in the zero-temperature vacuum case, the flat space (\ref{flatspace}) is not the 
unique AF instanton; there exist other 
periodic solutions of the Euclidean vacuum Einstein equations. A familiar instanton $g^S_{ab}(x)$ 
is the Euclidean section of the Schwarzschid solution \cite{gibbons0} :
\begin{equation} 
d^2s = \alpha d\tau^2 + \alpha^{-1}dr^2 +r^2d\theta^2+r^2\sin^2\theta d\phi^2;\quad \alpha=\left(1-\frac{2GM}{r}\right)^{1/2}
\label{schar}
\end{equation}
in Boyer-Lindquist coordinates $(t,r,\theta,\phi)$, where $M$ is a mass parameter. The metric field must be
periodic in $\tau$ with period $\beta$, as a result, $M$ is not a fixed mass, but rather determined 
by inverse temperature $\beta$ \cite{gross},
\begin{equation}
\beta=8\pi GM.
\label{masstem}
\end{equation}
This gravitational instanton (\ref{schar}) is special classical field configuration characterized 
by its topological properties: 
zero self-intersection number and an Euler number of two \cite{gibbons1,gibbons2}.
The classical Euclidean action of such an instanton is non-zero \cite{gross}, 
\begin{equation}
S_E(g^S)=\frac{1}{2}\beta M,
\label{inactionc}
\end{equation}
at finite inverse temperature $\beta$. This instanton action is not the contribution of a particle, or soliton, 
of mass $M/2$, since $M$ depends on the temperature.   

The Schwarzschid solution (\ref{schar}) is a special case of the Kerr instanton \cite{gibbons1},
which is the Euclidean section of the Kerr solution. It has been widely conjectured that the Kerr instanton 
is the unique AF instanton other than flat space (\ref{flatspace}) \cite{lapedes80}.

The classical action of two instantons is given by
\begin{equation}
S^{(2)}_E(g^S)=2S_E(g^S) - \beta \frac{GM^2}{(\Delta z + 2GM)},
\label{2inactionc}
\end{equation}
where the first term is the action of two free instantons and the second term is their Coulomb interaction,
and $\Delta z > 2GM$ is the distance between two instantons.
This Coulomb interaction vanishes for a large separation of two instantons ($\Delta z \gg 2GM$) and 
its maximum value for $\Delta z \rightarrow 2GM$ is four times smaller than the action of two free instantons. 

Based on the assumptions that the spatial distribution of ${\mathcal N}$ instantons is homogeneous and
the average distance between two instantons is $\langle\Delta z\rangle >2GM$, up to the nearest neighborhood Coulomb
interactions, the total classical Euclidean action is approximately given by,  
\begin{equation}
S^{({\mathcal N})}_E(g^S)\approx {\mathcal N}S_E(g^S)\left[1 -  \frac{GM}{(\langle\Delta z\rangle + 2GM)}\right],
\label{ninactionc}
\end{equation}
where ${\mathcal N}S_E(g^S)$ is the classical action of ${\mathcal N}$ free instantons. In the following, we
will neglect the Coulomb interaction and adopt the approximation,
\begin{equation}
S^{({\mathcal N})}_E(g^S)\approx {\mathcal N}S_E(g^S),
\label{freein}
\end{equation} 
since the correction due to the Coulomb interaction, i.e. the 
second term in Eq. (\ref{ninactionc}), is less than $25\%$.

\subsection{Vacuum to vacuum transition amplitude}

The functional integral $Z$ (\ref{function0}) represents the transition amplitude from vacuum to 
vacuum. The canonical partition function $Z(g^S)$ (\ref{functionalf}) is proportional to the 
transition amplitude from the vacuum of zero-instanton ($\delta_{ab}$ flat space) 
to the vacuum of one-instanton $g^S$ at the finite temperature $T$. This transition amplitude is 
exponentially weighted by the finite classical action $S_E(g^S)$ of the instanton,
\begin{equation}
Z^{(1)}(g^S)=
\frac{\langle g^S_{ab},0 |\delta_{ab}, 0 \rangle_{\rm T\not=0}}{\langle \delta_{ab},0 |\delta_{ab}, 0 \rangle_{\rm T=0}} 
\propto e^{-S_E(g^S)}=e^{-{1\over2}\beta M},
\label{inz1}
\end{equation}
where the amplitude is normalized by the amplitude in flat spacetime and at zero temperature 
[see Sect. (\ref{zerosec})].
Whereas the transition amplitude from the vacuum of zero-instanton to the vacuum of 
${\mathcal N}$-instanton sectors with different topological numbers is, 
\begin{equation}
Z^{({\mathcal N})}(g^S)=
\frac{\langle  {\mathcal N}g^S_{ab},0 |\delta_{ab}, 0 \rangle_{\rm T\not=0}}{\langle \delta_{ab},0 |\delta_{ab}, 0 \rangle_{\rm T=0}}
\propto e^{-{\mathcal N}S_E(g^S)}=e^{-{1\over2}\beta M {\mathcal N}},
\label{inzn}
\end{equation}
where the Coulomb interaction between instantons is neglected. The transition amplitudes (\ref{inz1}) and (\ref{inzn})
indicates the configurations of instantons are weighted by $e^{-{1\over2}\beta M {\mathcal N}}$ in the functional 
integral (\ref{einteval}) and (\ref{functionalf}). 

This is reminiscent of the t'Hooft gauge instanton. In the Euclidean space, the vacuum-to-vacuum amplitude
of quantum gauge fields about the classical solution \cite{belavin1975} is exponentially weighted by 
the action ($8\pi^2/g^2$) of the t'Hooft gauge instanton \cite{thooft1976}, where $g$ is a gauge coupling. 
This gives the transition amplitude of gauge instantons with different winding numbers. These studies have led to the $\theta$- vacuum and resolution of the $U(1)$ problem. 

When a thermal equilibrium is reached for a large
Euclidean time interval $\Delta \tau$, and summing over all contributions of instanton sectors, 
the amplitude of the vacuum to vacuum transition (\ref{inzn}) has the form
\begin{equation}
\exp \left[(\Delta {\mathcal E}+i\frac{\Gamma}{2})\Delta \tau  V\right],
\label{inznrate}
\end{equation} 
where $\Delta {\mathcal E}$ ($\Gamma$) is a transition energy (rate) per unit of volume $V$. This 
relats to the real (imaginary) part of 
free energy ${\mathcal F}$ (\ref{functionalf}) or effective action $S^{\rm eff}_E(g^s)$ (\ref{effeuclidean}).  
In next section, we will discuss 
that such vacuum transition is interpreted as the decay of the flat-space vacuum at a finite temperature,
via the thermal nucleation of black holes.  

\section{Instability of flat space at finite temperature}\label{instabilitysec}

\subsection{Instability of flat space}

The quantum (thermal) fluctuations $\phi_{ab}$ about the Schwarzschid instanton $g^S$ 
[see Eq. (\ref{fluctuations})]
are determined by the eigenvalues and eigenfunctions of the operator $[{\mathcal M}(g^S)]_\beta$ 
[see Eq. (\ref{functionalf})]. 
It was found \cite{page,gross,unstable} that in addition to semi-positive eigenvalues, 
the operator ${\mathcal M}(g^S)_\beta$ 
has a negative eigenvalue $\lambda\simeq -0.19 (GM)^{-2}$, and corresponding mode $\phi^\lambda_{ab}$
of quantum (thermal) fluctuation. The functional integral (\ref{functionalf}) indicates that $\phi^\lambda_{ab}$ 
is an unstable mode for small quantum (thermal) fluctuations about the the Schwarzschid instanton $g^S$. 
This instanton is therefore not a strict minimum, rather an unstable saddle point of the action. 
Its role in the thermodynamics of hot gravitons is similar to the top of
the potential barrier. The rolling off the top of the potential barrier is caused by
an infinitesimal perturbation. It was shown that the infinitesimal quantum (thermal) fluctuation
$\epsilon\phi^\lambda_{ab}$ about the Schwarzschid instanton $g^S$,
\begin{equation}
g_{ab}(x)=g_{ab}^S(x)+\epsilon\phi^\lambda_{ab}(x),
\label{fluctuations1}
\end{equation}
decreases the Euclidean classical action $S_E(g^S)$ (\ref{inactionc}) by \cite{gross}
\begin{equation}
\delta S_E(g^S)=- 9.4\cdot 10^{-4}\epsilon^2 (GM)^{-2}.
\label{deactionc}
\end{equation}
Consequently the functional integral (\ref{functionalf}) is actually divergent,
indicating the instability of running away from the classical action. 

\begin{figure}[th]
\begin{center}
\includegraphics[]{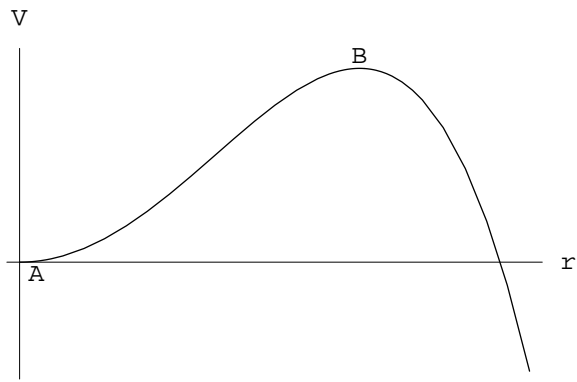}
\end{center}
\caption{Instability of flat spacetime at finite temperature. $V$ is the energy potential of Euclidean action.
$V(A)$ is the minimum of energy-potential and $V(B)$ is the maximum of energy potential.}%
\label{potential}%
\end{figure}

This instability can be illustrated by Fig. (\ref{potential}), where $V$ is the energy potential of the Euclidean action. At finite temperature $T$, the Euclidean action has two extremes:(classical solutions of equation of motion): one represented by $A$ that is flat space solution $\delta_{ab}$; another represented by $B$ that the Schwarzschid instanton solution $g_{ab}^S$. At a very low temperature 
$T\gtrsim V(A)\approx 0$, fluctuations about the state $A$ of the flat space are not stable, 
due to both the classical Jean instability of long-wave length fluctuations and quantum mechanical tunneling of low-energy fluctuations through the potential barrier. While, at very high temperature, thermal fluctuations are dominate. The states at temperature $T > V(B)$ degenerate to the state $B$ of the Schwarzschid instanton $g_{ab}^S$
at temperature $T= V(B)$, which simply sits at the top $B$ of the potential barrier. These states are highly unstable,  
due to thermal fluctuations over the potential barrier decrease Euclidean action. As results, free energy decreases, and
these states decays by gaining energy, i.e., rolling down from the top of energy potential.

\subsection{Decay rate of flat space via thermal nucleation}

In Ref. \cite{gross}, this instability is interpreted as a decay of flat space (\ref{flatspace}) 
at finite temperature $T$, 
and the decay process proceeds by quantum (thermal) fluctuations spontaneously 
nucleating black holes of radius $R=(4\pi T)^{-1}$ and mass $M=(8\pi GT)^{-1}$. For all temperatures,
the decay is only given by the saddle point at the top of the action, 
not given by the quantum tunneling process.

This instability gives rise to an imaginary part in the free energy ${\mathcal F}(g^S)$ (\ref{functionalf}), 
which is related to the rate $\Gamma$ (\ref{inznrate}) of decay transition from the zero-instanton (flat) 
vacuum to the  ${\mathcal N}$-instanton vacuum. Since the decay process proceeds by nucleating black holes,
this decay rate $\Gamma$ is interpreted as the rate of thermal nucleating black holes per unit of volume,
\begin{equation}
\Gamma \equiv \frac{d^4{\mathcal N}}{(g)^{1/2}d\tau d^3x },
\label{gammad}
\end{equation} 
at a finite temperature $T$.
In Refs. \cite{affleck,gross}, the thermal nucleation rate 
$\Gamma$ (\ref{gammad}) is given by,
\begin{equation}
\Gamma =\frac{\omega_0\beta}{\pi}{\rm Im}\left(\frac{\mathcal F}{V}\right)
\label{rateaff}
\end{equation}
where $\omega_0$ is determined by the negative eigenvalue $\lambda$,
\begin{equation}
\omega_0^2 =-\lambda \simeq \left(\frac{1.74}{\beta}\right)^2,
\label{rateomega}
\end{equation}
for all temperatures. 

To obtain the thermal nucleation rate $\Gamma$, one has to calculate the functional integral 
$Z$ (\ref{function0}) that integrates over the contributions from all instanton sectors ${\mathcal N}$. 
In Ref. \cite{gross}, the canonical partition function (\ref{functionalf}) for one instanton sector was calculated,
\begin{eqnarray}
Z^{(1)}(g^S)= \frac{i}{2} (\mu\beta)^{212/45}\left[\frac {M}{2\pi\beta}\right]^{3/2}VZ^0 \exp\left[-S_E(g^S)\right],
\label{partion1}
\end{eqnarray}
where the imaginary factor of $i/2$ occurs from the one normalizable negative mode $\phi^\lambda_{ab}$, $Z^0$ 
(\ref{zfree0}) is the contributions from the stable quantum (thermal) fluctuations about the instanton, 
the factor $(\mu\beta)^{212/45}$ arises from the renormalization counterterm and the regular mass $\mu$ should 
be taken of order the Planck mass $m_p$. Neglecting the classical Coulomb interactions (\ref{inactionc})
between the instantons, one estimated the contribution from the ${\mathcal N}$-instanton sector, 
\begin{eqnarray}
Z^{({\mathcal N})}(g^S)= \left[\frac{i}{2}\right]^{\mathcal N} (\mu\beta)^{212{\mathcal N}/45}
\frac{1}{{\mathcal N}!}\left[\frac {V}{64\pi^3G^{3/2}}\right]^{\mathcal N} Z^0 \exp\left[-{\mathcal N}S_E(g^S)\right].
\label{partionn}
\end{eqnarray}
Summing over all instanton sectors, one obtained
\begin{eqnarray}
Z(g^S)&=&\sum_{{\mathcal N}=0}^\infty Z^{({\mathcal N})}(g^S)\nonumber\\
&\simeq & Z^0 \exp\left\{\frac{i}{2}(\mu\beta)^{212/45}
\left(\frac {V}{64\pi^3G^{3/2}}\right) \exp\left[- S_E(g^S)\right]\right\},
\label{partionall}
\end{eqnarray}  
and the free energy ${\mathcal F}=\beta^{-1}\ln Z$.

The thermal nucleation rate $\Gamma$ per unit volume 
is thus given by Eqs. (\ref{rateaff},\ref{partionall}),  
\begin{eqnarray}
\Gamma\simeq \frac{0.87}{\beta}(\mu\beta)^{212/45} \left(\frac{m_p^3}{64 \pi^3}\right)
\exp\left[-S_E(g^S)\right].
\label{inrate0}
\end{eqnarray}
This formula gives the nucleation rate for pure gravity. It was extended to a general theory containing 
any number of massless matter field \cite{rate},
\begin{eqnarray}
\Gamma &=& 0.87 ({m_p\over T})^{\theta -1} {m_p^4\over 64 \pi^3}
\exp\left(-{m_p^2\over 16 \pi T^2}\right),
\label{inrate}\\
\theta &=&{1\over 45}\left(212n_2-{233\over 4}n_{3/2}-13n_1+{7\over 4}n_{1/2}+n_0\right),
\label{theta}
\end{eqnarray}
where $n_s$ is the number of massless spin-$s$ fields, the regular mass $\mu\simeq m_p$ and the classical
action $S_E(g^S)=m_p^2/(16 \pi T^2)$. 
The thermal nucleation rate (\ref{inrate}) reaches its maximum at high-temperature $T\sim m_p$, 
and exponentially suppressed at low-temperature $T\ll m_p$. This indicates that thermal nucleation process 
could be very important in the very early Universe.

Once these black holes have been nucleated, the unstable mode of small quantum (thermal) fluctuation
$\epsilon\phi^\lambda_{ab}$ will correspond to the subsequent
expansion (or collapse) of the black hole as it absorbs (or emits) thermal radiation. Thus these
black holes are colloquially termed quantum black holes in this article. The variations
of the quantum black hole mass $M$ and area $A=16\pi (GM)^2$ are given by \cite{gross}
\begin{eqnarray}
M&+&\delta M,\quad \delta M= 9.4\cdot 10^{-3} \epsilon(GM)^{-1};\label{variationm} \\
A&+&\delta A,\quad \delta A= 9.4\cdot 10^{-1} \epsilon G ,\label{variationa}
\end{eqnarray}
where positive and negative values of $\epsilon \sim M^2G$ correspond respectively 
expansion and collapse of the quantum black holes. These quantum black holes absorb (or emit) thermal radiation, 
then expand (or collapse), depending on the temperature of these quantum black holes 
$T=1/(8\pi GM)$ being larger (or less) than the temperature of their environment. Henceforth, we adopt the 
notation QBH standing for quantum black hole.

\section{Gravitational instantons and cosmological term}

\subsection{The flat Robertson-Walker metric}

Suppose that the space-time symmetry of Universe is described by the Robertson-Walker line element 
with zero curvature $k=0$ and the scale factor $a(\tau)$,
\begin{equation}
d^2s = d\tau^2 + a^2(\tau) (dx^2+dy^2+dz^2)=a^2(\eta)\left[d\eta^2 + dx^2+dy^2+dz^2\right],
\label{robertson} 
\end{equation}
where the Euclidean time interval $d\tau=-idt$, and conformal time interval $d\eta=d\tau/a(\tau)$. 
The Robertson-Walker metric $g^R_{ab}(x)=a(\tau)\delta_{ab}$, which is a space-time geometry 
that differs from the flat space-time geometry only by an overall scale factor $a(\tau)\ge 0$. 

Suppose that Universe is described by a canonical ensemble at temperature $T=1/\beta$, 
the Robertson-Walker metrics $g^R_{ab}$ in Eq. (\ref{robertson}) is strictly periodic 
in Euclidean time $\tau$ with period $\beta$: $g^R_{ab}(\tau,x)=g^R_{ab}(\tau+\beta,x)$.  
The metric $g^R_{ab}(\eta,x)=g^R_{ab}(\eta+1,x)$, is strictly periodic
in the Euclidean conformal time $\eta$ with period $1$, where we use the entropy conservation $aT=1$ and 
temperature $T$ is in unit of the Planck mass $m_p$.

The flat Robertson-Walker metrics $g^R_{ab}$ (\ref{robertson}) is a classical solution to 
the vacuum Einstein equation (\ref{e5}) and the Ricci scalar $R=0$, provided,
\begin{equation}
\frac {3}{a^2} \dot a^2(\tau)=0,\quad \dot a=\frac {da}{d\eta}=a\frac {da}{d\tau}.
\label{friedmann} 
\end{equation}
The scale factor $a$ is a time-independent constant, at a fixed temperature $T$. Thus 
the flat Robertson-Walker metrics $g^R_{ab}=a\delta_{ab}$, is a trivial AF instanton, 
which has zero action $S_E(g^R)=0$. The classical solution $g^R$ is a strictly local 
minimum of the action ($A$ in Fig. (\ref{potential})). 
The discussions on this local minimum can be found in section (\ref{flatnonzero}).  

\subsection{Finite action of ${\mathcal N}$ instantons in functional integral}
 
As discussed in previous sections (\ref{finitesec}-\ref{instabilitysec}), there is another 
non-trivial AF Schwarzschid instanton (\ref{schar}) with the finite action $S_E(g^S)$ (\ref{inactionc}) 
at temperature $T$. There is a topologically distinct configuration of ${\mathcal N}$-instanton sector 
whose finite action is approximately given by the finite action of ${\mathcal N}$ free instantons,
i.e., ${\mathcal N}S_E(g^S)$ [see Eq. (\ref{ninactionc})]. 
The flat Robertson-Walker metrics $g^R_{ab}=a\delta_{ab}$ at temperature $T$
is not stable, against quantum (thermal) fluctuations. This flat Robertson-Walker spacetime is bound to 
spontaneously decay,
via the thermal nucleation of QBHs. The rate of thermal nucleation process is given by 
Eq. (\ref{inrate}). This is an energy-gain process by decreasing finite action.

Since the instanton action ${\mathcal N}S_E(g^S)$ is finite and the thermal nucleation rate is nonzero, 
these topologically distinct configurations of instantons are not vanishing, and must be included into
configurations of functional integral (\ref{function0}) and (\ref{functionalf}) \cite{gross}. 
This is the same as the reason for including 
t'Hooft instantons into the functional integral in non-Abelian gauge theories \cite{weinbergbookfield}.
Thus, functional integral (\ref{functionalf}) for quantum gravity at finite temperature $T=\beta^{-1}$ 
becomes,
\begin{equation}
Z(\beta)=\int_{\rm instantons} {\mathcal D}[g_{ab}(x)]e^{-{\mathcal N}S_E(g^S)-S_E(g)+ {\rm gauge-fixing}\hskip0.05cm {\rm terms}},
\label{functionali}
\end{equation}
where functional integral is carried over ${\mathcal N}$-instanton configurations. The Euclidean action $S_E(g)$ is 
given in Eq. (\ref{eactiont}). This indicates that the effective action for quantum gravitational field at finite temperature contains contributions from gravitational instantons. 

As discussed in section (\ref{instabilitysec}), that the finite action of ${\mathcal N}$-instanton sector in 
functional integral (\ref{functionali}) indicates the decay process by thermal nucleating QBHs.
The decay rate $\Gamma$ at a finite temperature $T$ is given by Eqs. (\ref{gammad},\ref{inrate0},\ref{inrate}),
independently from space ${\bf x}$ and time $\tau$. 
Therefore, the total number of QBHs nucleated ${\mathcal N}$ is given by 
\begin{equation}
{\mathcal N} = \Gamma V\Delta\tau,  
\label{numberfix}
\end{equation}
where $\Delta \tau$ is the Euclidean time interval when thermal nucleation 
process takes place. The finite action of 
${\mathcal N}$-instantons in the canonical partition function (\ref{functionali}) is then,
\begin{equation}
{\mathcal N}S_E(g^S)=  \left(\frac{1}{2}\beta M \right)\Gamma V\Delta\tau,
\label{faction}
\end{equation}
where the finite action for a single instanton $S_E(g^S)=\beta M /2$ (\ref{schar}) is time-independent.

\subsection{Partition function for time-evolution of canonical ensemble}

So far we have discussed the functional integral (partition function) describing a canonical ensemble at 
a fixed temperature. Now we turn to study the functional integral describing the adiabatic evolution of 
a canonical ensemble, where temperature $T=\beta^{-1}$ is a function of time. We consider 
a sequence of canonical ensembles 
${\mathcal S}(\tau_0),{\mathcal S}(\tau_1),{\mathcal S}(\tau _2),{\mathcal S}(\tau _3),
\cdot\cdot\cdot,{\mathcal S}(\tau _N)$, 
at different temperatures $T(\tau_0)>T(\tau_1)>T(\tau _2)>T(\tau _3)>\cdot\cdot\cdot>T(\tau _N)$, and
scaling factors $a(\tau_0)<a(\tau_1)<a(\tau _2)<a(\tau _3)<\cdot\cdot\cdot<a(\tau _N)$, at different times  
$\tau_0<\tau_1<\tau _2<\tau _3<\cdot\cdot\cdot<\tau _N$, in comoving frame. 
In this case, both the mass parameter
$M(\tau_i)$ (\ref{masstem}) and thermal nucleation rate $\Gamma(\tau_i)$ (\ref{inrate}) of QBHs
are functions of time 
$\tau_i$. 
At each instantaneous moment $\tau_i$ and temperature $T(\tau _i)$, the 
canonical partition function is given by the functional integral (\ref{einteval},\ref{eactiont},\ref{functionalf}) and (\ref{functionali}) with normalization $N_\beta=1$.
For an infinitesimal time interval $d\tau=\tau_{i+1}-\tau_i$, we assume that 
the time-evolution of canonical ensemble ${\mathcal S}(\tau_i)$ is strictly adiabatic and 
total entropy of canonical ensemble is conserved, i.e., $a(\tau_i)=\beta(\tau_i)$.
  
We define the functional integral for the time-evolution of canonical ensembles as a time-ordering 
product of canonical partition functions (\ref{functionali}) for canonical ensembles at 
different temperature $T(\tau_i)=\beta^{-1}(\tau_i)$,
\begin{equation}
{\mathcal Z}={\mathcal T}\prod_{i=0}^NZ[\beta(\tau_i)]\equiv Z[\beta(\tau_0)]\cdot Z[\beta(\tau_1)]\cdot Z[\beta(\tau_2)]
\cdot\cdot\cdot Z[\beta(\tau_N)].
\label{tfunctionali}
\end{equation}
Further, we define the integration over Euclidean time 
sequence $\tau_i$ as,
\begin{eqnarray}
\int_{\tau_0}^{\tau} d\tau ' \equiv \sum^N_{i=0} \int_0^{\beta(\tau_i)} d\tau '=\sum^N_{i=0} \beta(\tau_i),
\quad \tau\equiv \tau_N,
\label{timeeinteval}
\end{eqnarray}
four-dimensional volume element as,
\begin{eqnarray}
\int (g)^{1/2}d^4x=\sum_{i=0}^N \beta(\tau_i)V(\tau_i) = \sum^N_{i=0}\int_0^{\beta(\tau_i)} 
d\tau '\int^{V(\tau_i)} (g)^{1/2}d^3x,
\label{voleinteval}
\end{eqnarray}
and the measure of functional integral over different time period $\beta(\tau_i)$ as,
\begin{eqnarray}
\int {\mathcal D}[g_{ab}(x)]
\equiv\prod_{i=0}^N\prod_{\tau '=0}^{\beta(\tau_i)}\int {\mathcal D}[g_{ab}({\bf x},\tau ')].
\label{timeeintevalm}
\end{eqnarray}  
Then, the functional integral (\ref{tfunctionali}) can be written as,
\begin{eqnarray}
{\mathcal Z}=\prod_{i=0}^N\prod_{\tau '=0}^{\beta(\tau_i)}\int {\mathcal D}[g_{ab}({\bf x},\tau ')]
e^{-{\mathcal N}S_E(g^S)-S_E(g)+ {\rm gauge-fixing}},
\label{tfunctionalf}
\end{eqnarray}
where the Euclidean action $S_E(g)$ and finite instanton action ${\mathcal N}S_E(g^S)$ are respectively given by 
the sum over their counterparts at different temperature $T(\tau_i)$. The Euclidean action $S_E(g)$ is,
\begin{eqnarray}
S_E(g)=-\frac{1}{16\pi G} \sum_{i=0}^N \int_0^{\beta(\tau_i)} d\tau '\int (g)^{1/2}d^3x R + {\rm boundary}\hskip0.15cm {\rm terms}.
\label{timeeactiont}
\end{eqnarray}
The finite ${\mathcal N}$-instanton action ${\mathcal N}S_E(g^S)$ (\ref{faction}) becomes,
\begin{eqnarray}
{\mathcal N}S_E(g^S)=\sum_{i=0}^N \beta(\tau_i)V(\tau_i) \left[\Delta\tau_i \frac{1}{2} M(\tau_i)\Gamma(\tau_i)\right].
\label{timeinactionc1}
\end{eqnarray}
The factor $\sum_{i=0}^N \beta(\tau_i)V(\tau_i)$ gives the four dimensional volume (\ref{voleinteval}) in the 
function integral (\ref{timeeintevalm},\ref{tfunctionalf}). Because the mass parameter $M$ and decay rate
$\Gamma$ in Eq. (\ref{timeinactionc1}) are time-dependent, we should integrate $M(\tau_i)\Gamma(\tau_i)/2$ 
over the time-interval $\Delta \tau_i=\tau_i-\tau_0$ when thermal nucleating process occurs,  
\begin{eqnarray}
\Delta\tau_i \frac{1}{2} M(\tau_i)\Gamma(\tau_i)\rightarrow
\int_{\tau_0}^{\tau_i}d\tau ' \frac{1}{2} M(\tau ')\Gamma(\tau ').
\label{timeinactionc2}
\end{eqnarray}
As results, the finite ${\mathcal N}$-instanton action ${\mathcal N}S_E(g^S)$ (\ref{timeinactionc1}) is given by
\begin{eqnarray}
{\mathcal N}S_E(g^S)=\int (g)^{1/2}d^4x\rho_\Lambda(\tau_0,\tau),
\label{timeinactionc3}
\end{eqnarray}
where
\begin{eqnarray}
\rho_\Lambda(\tau_0,\tau)\equiv \int_{\tau_0}^{\tau}d\tau ' \frac{1}{2} M(\tau ')\Gamma(\tau ').
\label{drho}
\end{eqnarray}  
Actually, $\rho_\Lambda(\tau_0,\tau)$ is the energy-density of ${\mathcal N}$ QBHs nucleated in the Euclidean time interval $\Delta=\tau-\tau_0$.

\subsection{Minkowski formulation}

Performing the back Wick rotation from the Euclidean signature $(++++)$ to Minkowski one 
$(-+++)$, we can obtain the effective action and functional integral in the Minkowski formulation. 
The Euclidean action (\ref{timeeactiont}) becomes the Minkowski one,
\begin{eqnarray}
S(g)=-\frac{1}{16\pi G} \int (-g)^{1/2}d^4x R + {\rm boundary}\hskip0.15cm {\rm terms},
\label{mtimeeactiont}
\end{eqnarray}
where $(-g)^{1/2}d^4x$ is the Minkowski volume element, and functional integral measure (\ref{timeeintevalm}) 
is changed correspondingly. 
The finite action of ${\mathcal N}$-instanton actions (\ref{timeinactionc3}) becomes the cosmological term, we write
it as
\begin{eqnarray}
S_\Lambda(g^S)=\int (-g)^{1/2}d^4x\rho_\Lambda(t_0,t),
\label{mtimeinactionc1}
\end{eqnarray}
where 
\begin{eqnarray}
\rho_\Lambda(t_0,t)\equiv\int_{t_0}^{t}dt ' \frac{1}{2} M(t ')\Gamma(t '),
\label{mdrho}
\end{eqnarray}
and $\Gamma(t')$ is the thermal nucleation rate per unit of Minkowski volume [see Eq. (\ref{gammad})] 
\begin{equation}
\Gamma \equiv \frac{d^4{\mathcal N}}{(-g)^{1/2}d td^3x }.
\label{mgammad}
\end{equation}
The cosmological constant is then given by,
\begin{eqnarray}
\Lambda=8\pi G\rho_\Lambda(t_0,t).
\label{lambdarho}
\end{eqnarray} 

The above formulations and discussions can be straightforwardly extended 
to the case that matter fields are present. The total effective action in functional integral is then,
\begin{eqnarray}
S(g)+S_\Lambda(g^S)+S_M,
\label{totala}
\end{eqnarray}
where $S_M$ is the action for fundamental particles, whose energy momentum tensor is $T_{ab}$. In fact,
the non-zero temperature $T$ is attributed to the thermal energy of relativistic particles.  
Based on the principle of least action, we obtain the Einstein equation (\ref{e1}) 
with the cosmological term given by Eq. (\ref{mtimeinactionc1}). Such cosmological term has two features: (i) its geometric origin from the nucleation of gravitational instantons, and appearance in the right-handed side of Einstein equation (\ref{e1}); (ii) its energetic origin from the energy-gain of nucleating QBHs 
$\delta E_\Lambda=-\delta E$, where $\delta E<0$ is the vacuum-energy variation of the flat space at finite temperature. 
This temperature is due to thermal energy of relativistic particles. 
The energy-momentum tensor corresponding the cosmological term is $T^\Lambda_{ab}=-g_{ab}\rho_\Lambda$. 
The total energy conservation is given by $(T_{ab}+T^\Lambda_{ab})^{;a}=0$ in the process of nucleating QBHs.

\section{Early evolution of Universe}

\subsection{Inflation}

We consider the post-Planckian era of very early Universe, at the time $t\geq t_0=1$, the scale factor 
$a(t_0)>a(t_0)=1$ and the temperature $T\leq T_0=1$, where and henceforth, the temperature $T$ and time $t$ are in the Planck unit.
The temperature is due to the thermal graviton gas with energy-density $\rho_0$ (\ref{energy0}).
Suppose that the early Universe is given by the spatial-flat geometry of 
the Robertson-Walker line element with the scale factor $a(t)$ and zero curvature $k=0$ (\ref{robertson}), 
and the Einstein equation (\ref{e1}) describing Universe expansion becomes,
\begin{equation}
H^2\equiv \left({\dot a\over a}\right)^2 = {8\pi G \over 3}\left[\rho_0(t) + \rho_\Lambda(t)\right].
\label{requation} 
\end{equation}
where $\rho_\Lambda(t)$ 
is given by Eq. (\ref{mdrho}) with the initial time $t_0=1$ and temperature $T_0=1$.
\begin{eqnarray}
\rho_\Lambda(t)&=&{1\over2}\int_{t_0}^t d\tau M(\tau)\Gamma [T(\tau)],
\nonumber\\
&=& {0.87m_p^4\over 1024\pi^4}\int_{t_0}^td\tau 
\left({1\over T(\tau)}\right)^\theta e^{-{1\over 16\pi T^2(\tau)}} 
\label{rho}
\end{eqnarray}
which is the energy-density of QBHs nucleated from initial time $t_0$ 
and temperature $T_0$ to final time $t$ and temperature $T(t)$. Corresponding to the QBHs' 
energy-density (\ref{rho}), the QBHs' energy-density is given by,
\begin{eqnarray}
n_\Lambda(t)&=&{1\over2}\int_{t_0}^t d\tau \Gamma [T(\tau)],
\nonumber\\
&=& {0.87m_p^3\over 128\pi^3}\int_{t_0}^td\tau 
\left({1\over T(\tau)}\right)^{\theta-1} e^{-{1\over 16\pi T^2(\tau)}}. 
\label{nrho}
\end{eqnarray}

The Universe expands and its temperature $T(t)$ decreases, $a(t)T(t)=1$ for 
the entropy-conservation. As results, the energy-density $\rho_0(t)$ (\ref{energy0})
decreases, whereas $\rho_\Lambda(t)$ (\ref{rho}) increases and 
asymptotically approaches a constant $\bar\rho_\Lambda$ for $t\gg 1$. 
Thus, Eq.(\ref{requation}) implies an inflationary Universe. How does such inflation phase end? 

\subsection{\it Hawking radiation and reheating.}

When the Universe temperature $T(t)$ is smaller than the temperature 
$T(\tau)=1/(8\pi GM(\tau))$ of QBHs that are created at an 
earlier time $\tau <t$, i.e., $T(t)<T(\tau)$, these QBHs lose their masses by the 
Hawking radiation. On the other hand, accretion occurs if $T(t)>T(\tau)$ (see Eqs. (\ref{variationm}) 
and (\ref{variationa})). 
If we were only to consider emission and absorption of gravitons, 
the mass-variation of QBHs is given by \cite{hawking,accretion} 
\begin{equation}
{\delta M(\tau)\over \delta\tau} = {\pi^2\over 15}[T^4(t)-T^4(\tau)]4\pi R^2(\tau), 
\label{hrate1} 
\end{equation}
where the black hole size $R(\tau)=2M(\tau)$. For $T(\tau)\gg T(t)$, we approximately
obtain,
\begin{equation}
M_H(t)\simeq M(\tau)[1-{2\pi^2\over 5}T^3(\tau)(t-\tau)]^{1/3}, 
\label{hrate2} 
\end{equation}
indicating that QBHs lose their mass-energy. We speculate that the extremal minimum of QBHs'
mass-energy is the order of the Planck scale.   
 
The Hawking process (\ref{hrate2}) occurs, contemporaneously with QBH nucleation. Due to the 
Hawking radiation, the energy-density $\rho_\Lambda(t)$ (\ref{rho}) 
of QBHs is reduced. The variation of the energy-density $\rho_\Lambda(t)$ can be obtained by
Eq. (\ref{rho}),
\begin{eqnarray}
\Delta\rho_\Lambda(t) &=&{1\over2}\!\int_{t_0}^t d\tau \Delta M\Gamma [T(\tau)],
\label{deltarho}\\
&=& {0.87m_p^4\over 128\pi^3}\int_{t_0}^td\tau \Delta M
\left({1\over T(\tau)}\right)^{(\theta -1)}e^{-{1\over 16\pi T^2(\tau)}},\nonumber
\end{eqnarray}
where 
\begin{equation}
\Delta M = M_H(t) - M(\tau),
\label{deltam}
\end{equation}
and $M_H(t)$ is given by Eq. (\ref{hrate2}). The energy-density variation of QBHs $\Delta\rho_\Lambda(t)<0$. 
The energy-density $\rho_\Lambda(t)$ in Eq. (\ref{requation}) should be replaced by 
\begin{equation}
\rho(t)=\rho_\Lambda(t)+\Delta\rho_\Lambda(t)={1\over2}\int_{t_0}^t d\tau M_H(\tau)\Gamma [T(\tau)].
\label{totalrho}
\end{equation}
This shows that the energy-gain $\rho_\Lambda(t)$ in QBH nucleation is 
converted into the radiation energy $|\Delta\rho_\Lambda(t)|$, which starts to reheat the Universe.

Assuming $a(t)T(t)= 1$ and $\theta=212/45$ (gravitons), 
$t_0= 1$ and $T_0= 1$, we use the method of iterating procedure in the time
``$t$'' to numerically integrate Eqs. (\ref{requation},\ref{totalrho}) 
for the evolution of Universe, by taking into account 
both the nucleation process (\ref{inrate},\ref{rho}) and the Hawking process (\ref{hrate2},\ref{deltarho}). 
At the beginning ($t< O(10^2)$), the nucleation process dominates over the Hawking process.
The variation of energy-density $\rho(t)$ (\ref{totalrho}) is rather slow, so that the 
``graceful exit'' problem is avoided and the evolution of Universe is inflationary. However,
at the end ($t> O(10^2)$), the variation of energy-density $\rho(t)$ (\ref{totalrho}) is rather rapid, 
the Hawking process is dominant, eventually ends the inflation and reheats Universe.

\begin{figure}[th]
\begin{center}
\includegraphics[]{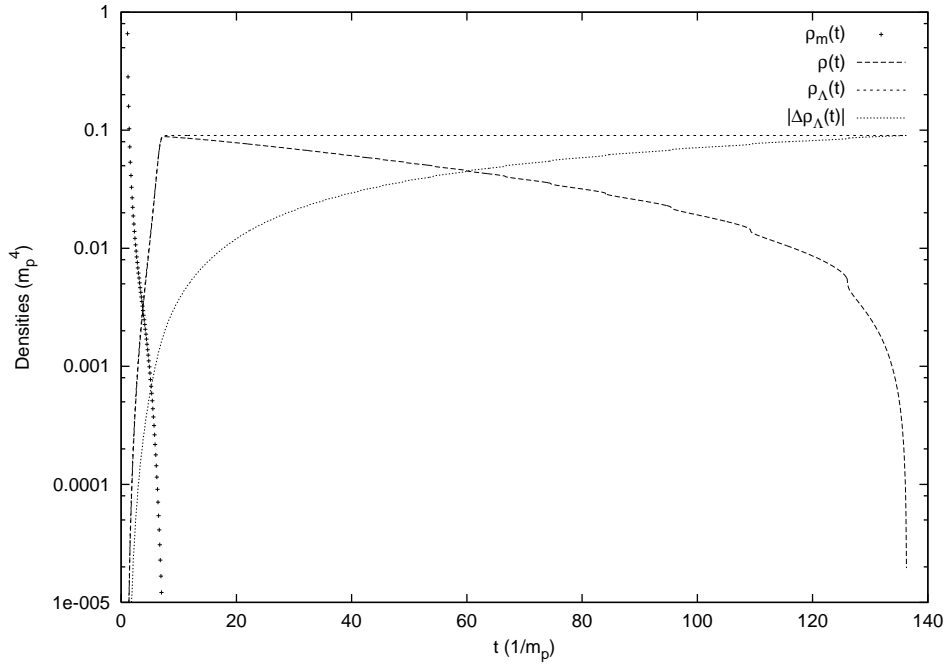}
\end{center}
\caption{The energy-densities $\rho_m(t)$, $\rho_\Lambda(t)$, $\rho(t)$ and 
$|\Delta\rho_\Lambda(t)|$ as functions of time.}%
\label{densities}%
\end{figure}

\subsection{Numerical results.}       

In the initial phase $t\sim 1$, the energy-density of QBHs 
$\rho_\Lambda(t)$ (\ref{rho}) is negligible, compared with the energy-density of
thermal graviton gas $\rho_0$ (\ref{energy0}). The solution to Eq. (\ref{requation}) is radiative,
\begin{equation}
a_1= a_0 t^{1/2},\hskip0.3cm T_1=T_0 t^{-1/2},
\label{radiative} 
\end{equation} 
where $a_0\sim 1$ and $T_0\sim 1$ are initial scaling-factor and temperature.
$a_0T_0\sim 1$ implies that the initial entropy $S_0=(a_0T_0)^3$ is given 
by $O(1)$ quantum states of Planck energy in the Planck volume.
As the time $t$ increases, $\rho_0(t)$ decreases, $\rho_\Lambda(t)$ increases 
and becomes dominant in Eq. (\ref{requation}). 

\begin{figure}[th]
\begin{center}
\includegraphics[]{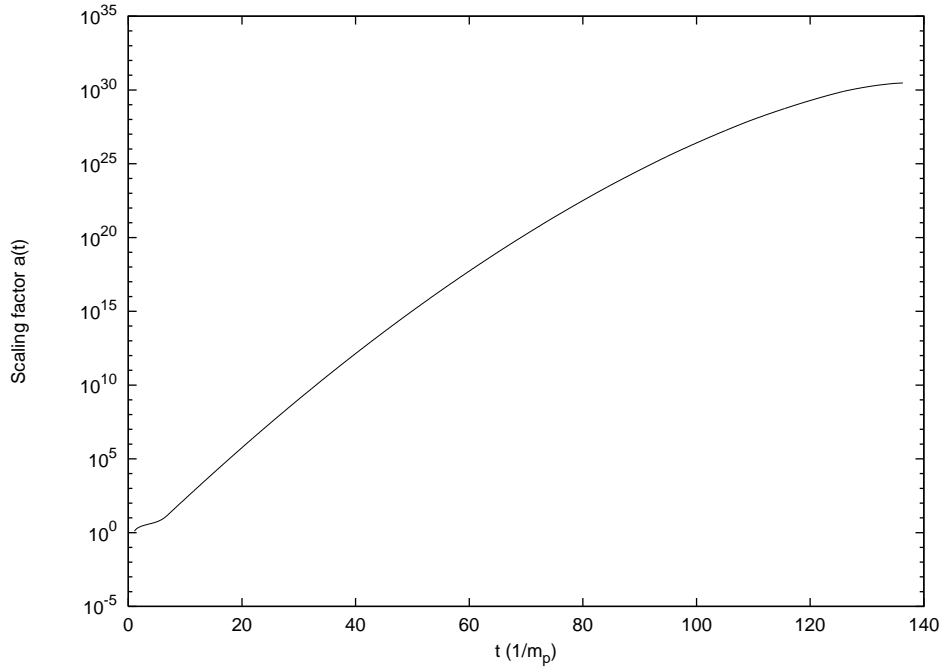}
\end{center}
\caption{The scaling factor $a(t)$ and temperature $T(t)=1/a(t)$ as functions of time.}%
\label{inflation}%
\end{figure}

In Fig. [\ref{densities}], we plot $\rho_0(t)$ (cross line) 
and $\rho_\Lambda(t)$ (short-dash line) as functions of the time $t$.
It is shown that (i) the pre-inflationary phase (\ref{radiative}) for $t<10$; (ii) the inflationary phase 
for $t>10$, where $\rho_0$ (\ref{energy0}) is vanishing and $\rho_\Lambda(t)$ is approaching to an 
asymptotic value 
\begin{equation}
\bar \rho_\Lambda\simeq 9.03\cdot 10^{-2}m_p^4,
\label{arho} 
\end{equation}
and the corresponding QBHs number-density (\ref{nrho}) is $\bar n_\Lambda\sim 10^{-2}m_p^3$.
In Fig. [\ref{densities}], 
we also plot the radiation energy $|\Delta\rho_\Lambda(t)|$ (\ref{deltarho}) (dot-line) 
and the energy-density $\rho(t)$ 
(\ref{totalrho}) (long-dash line), showing that $\rho(t)$ slowly varies for $8<t<100$ and 
$\rho(t)\rightarrow 0$ for $t>100$. Correspondingly, in Fig. [\ref{inflation}], we plot $a(t)$ 
and $T(t)\simeq 1/a(t)$, which show that (i) the pre-inflationary phase (\ref{radiative}) for $t<4$, 
(ii) an exponential inflation $a(t)\simeq a_0\exp (N_et)$ for $4<t<110$ and (iii) $a(t)$
approaching $10^{30}$ for $t>110$. 

We consider that the inflation ends at $t\simeq t_f$, when the QBHs' mass-energy 
(\ref{hrate2},\ref{totalrho}) is reduced to $M_H\simeq 1/(8\pi)$ 
corresponding to the QBH temperature $T_H\simeq 1$. 
We find that $t_f\simeq 113$, $a_f\simeq 2.84\cdot 10^{28}$ 
and the e-folding factor $N_et_f\simeq 65.5$. 

The reheating process occurs when $t\sim t_f$. We assume that apart from the Hawking radiation from each 
QBH, these QBHs collide each other and annihilate into other particles, since the energy-density 
$\rho(t)$ and number-density $n_\lambda$ of QBHs is still very high, comparable with Planck densities, 
at the end of inflation phase $t\simeq t_f$. 
We define that the reheating phase ends at time $t_h$, when $t\rightarrow t_h$, QBHs' densities 
$\rho(t)\rightarrow 0$ and $n_\lambda(t)\rightarrow 0$. 
In the other wards, the reheating phase ends at $t=t_h$, when the QBHs' energy-density $\bar \rho_\Lambda$ 
(\ref{arho}) has been completely converted into the energy-density of other relativistic particles, and QBHs has 
completely annihilated into other relativistic particles.
The reheating temperature $T_h$ at the end of reheating $t=t_h$ can be estimated by the thermal 
energy-density of thermal relativistic particles,
\begin{equation}
\rho_0^h=
g_s{\pi^2\over15}T_h^4=\bar \rho_\Lambda \simeq |\Delta\rho_\Lambda(t_h)|,
\label{reheatingt} 
\end{equation} 
where $g_s$ stands for the summation over contributions of all relativistic 
particles created in the Hawking process and QBHs annihilation. Since all possible relativistic particles 
are created, the reheating temperature must be smaller than the initial temperature, i.e., $T_h<T_0\simeq 1$. 
We are not able to determine $T_h$ and $t_h$, and 
leave them as parameters in this article. For $T_h\sim O(10^{-2})$, 
an enormous entropy $S_h= (T_ha_h)^3\sim O(10^{79})$ 
is produced. In the time interval $(t_f,t_h)$ for the reheating phase, 
$|\Delta\rho_\Lambda(t)|\rightarrow\bar \rho_\Lambda$, the scaling factor $a(t)$ slowly varies, see 
Fig. [\ref{inflation}].  

These numerical results Figs. [\ref{densities},\ref{inflation}] depend on the initial temperature $T_0$ and 
time $t_0$. In particular, the initial temperature $T_0$ plays a crucial role, for its 
exponential dependence $\exp -1/(16\pi T^2)$ in QBH nucleating rate (\ref{rho}). While, numerical results 
are not sensitive to the parameter $\theta$ (\ref{inflation}). 

\section {\it Cosmological constant at the present Universe.}

After reheating to the temperature $T_h$, the nucleation of QBHs starts 
again, and the density of energy-gain $\rho^h_\Lambda(t)$ is given by Eq. (\ref{rho})
with the initial time $t_h$ and temperature $T_h$. 
\begin{eqnarray}
\rho^h_\Lambda(t)&=&{1\over2}\int_{t_h}^t d\tau M(\tau)\Gamma [T(\tau)],
\nonumber\\
&=& {0.87m_p^4\over 1024\pi^4}\int_{t_h}^td\tau 
\left({1\over T(\tau)}\right)^\theta e^{-{1\over 16\pi T^2(\tau)}} ,
\label{rhoreh}
\end{eqnarray}
where temperature $T$ is smaller than the reheating temperature $T_h$ ($T<T_h$).
Because the reheating temperature $T_h$ is lower than the initial temperature $T_0$ ($T_h<T_0\simeq 1$),
QBHs' density $\rho^h_\Lambda(t)$ (\ref{rhoreh}) is 
exponentially suppressed by lower temperature $T\ll 1$, thus smaller than 
$\rho_0^h$ (\ref{reheatingt}) the thermal 
energy-density of thermal relativistic particles,
\begin{equation}
\rho^h_\Lambda\ll \rho_0^h\simeq T_h^4,\quad t>t_h.
\label{radiative0} 
\end{equation}
Therefore, the thermal energy-density $\rho_0^h$ dominates in Eq. (\ref{requation}) 
for the evolution of the Universe. As a consequence,
the Universe begins the evolution described by the Standard Cosmology:
\begin{equation}
a(t)= a_h (t/t_h)^\alpha,\hskip0.3cm T(t)=T_h (t/t_h)^{-\alpha},
\label{radiative1} 
\end{equation}
with total entropy $S_h=(a_hT_h)^3$. The index $\alpha$ is determined by radiation or matter dominate epochs.
We consider this as a new era initiated with $t_h,T_h$ and $a_h$,
independently from the inflationary era before the reheating.

The energy-density $\rho^h_\Lambda(t)$ is mainly contributed from QBHs nucleated 
in the reheating. Due to the Hawking radiation (\ref{hrate2}), the variation of energy-density 
$\Delta\rho^h_\Lambda(t)$ 
is given by Eq. (\ref{deltarho}) with initial time $t_h$ and temperature $T_h$,
\begin{eqnarray}
\Delta\rho^h_\Lambda(t) &=&{1\over2}\!\int_{t_h}^t d\tau \Delta M\Gamma [T(\tau)],
\label{deltarhoreh}
\end{eqnarray} 
and $\Delta\rho^h_\Lambda(t)<0$. Analogously to Eq. (\ref{totalrho}),  the total energy-density of 
QBHs nucleated after the reheating ear is, 
\begin{eqnarray}
\rho_h(t)&=& \rho^h_\Lambda(t) + \Delta\rho^h_\Lambda(t)\nonumber\\
&=&{0.87m_p^4\over 128\pi^3}\int_{t_h}^td\tau M_H
\left({1\over T(\tau)}\right)^{(\theta -1)}e^{-{1\over 16\pi T^2(\tau)}}.
\label{crhoreh} 
\end{eqnarray}
This is related to the ``dark-energy'' density, and the cosmological constant is given by 
\begin{equation}
\Lambda=8\pi G\rho_h(t).
\label{cosconst1} 
\end{equation}

Assuming that the mass of QBHs has been reduced to the minimal mass $M_H=1/(8\pi)$ 
at the present time $t\gg t_h$, we obtain, 
\begin{equation}
\rho_h(t)= {0.87m_p^4\over 1024\pi^4}\int_{t_h}^td\tau 
\left({1\over T(\tau)}\right)^{(\theta-1)} e^{-{1\over 16\pi T^2(\tau)}},
\label{crho1}
\end{equation} 
from Eq. (\ref{crhoreh}). Substituting solutions $a(t)$ and $T(t)$
(\ref{radiative1}) into $\rho_h(t)$ (\ref{crho1}), we obtain,
\begin{eqnarray}
\rho_h(t) &=& {0.87m_p^4\over 1024\pi^4}\int_{t_h}^td\tau 
\left({\tau^\alpha\over T_ht^\alpha_h}\right)^{(\theta-1)} 
e^{-{\tau^{2\alpha}\over 16\pi T^2_ht^{2\alpha}_h}} 
\nonumber\\
&=& {0.87m_p^4\over 1024\pi^4} {(16\pi)^\delta t_h\over 2\alpha }T_h^{1/\alpha}
\Gamma(\delta,z_2,z_1)\label{cosm2}
\end{eqnarray}
where the incomplete Gamma-function is given by
\begin{equation}
\Gamma(\delta,z_2,z_1) \equiv \int_{z_1}^{z_2}dx x^{\delta-1}e^{-x},
\label{gamma} 
\end{equation}
and
\begin{eqnarray}
\delta &=& \frac{\alpha(\theta-1)+1}{2\alpha};\label{gpara}\\
z_2&=& \frac{t^{2\alpha}}{16\pi T_h^2t^{2\alpha}_h},\quad
z_1= \frac{1}{16\pi T_h^2}. \nonumber
\end{eqnarray}
The asymptotic representation of the incomplete Gamma-function (\ref{gamma}),
\begin{equation}
\Gamma(\delta,z_2,z_1) \simeq z_1^{\delta -1}e^{-z_1}.
\label{gamma1} 
\end{equation}
for $z_2\gg z_1\gg 1$ and $z_2\rightarrow\infty$. 
Using this asymptotic representation, we obtain the ``dark-energy'' density and cosmological constant
at the present time $t\gg t_h$, 
\begin{equation}
\rho_h \simeq {0.87\over 64\pi^3} {m_p^4t_h\over 2\alpha}T_h^{(3-\theta)}
e^{-{1\over16\pi T_h^2}},\quad \Lambda=8\pi G \rho_h.
\label{cosm3} 
\end{equation}
which are approximately independently of time $t$.
Analogously to Eq. (\ref{nrho}), we compute the number density of QBHs created after the reheating phase,
\begin{equation}
n_h(t)= \int_{t_h}^td\tau \Gamma [T(\tau)] = {0.87m_p^3\over 128\pi^3}\int_{t_h}^td\tau 
\left({1\over T(\tau)}\right)^{(\theta-1)} e^{-{1\over 16\pi T^2(\tau)}},
\label{qbhn1}
\end{equation}
and obtain the number-density of QBHs at present time,
\begin{equation}
n_h \simeq {0.87\over 8\pi^2} {m_p^3t_h\over 2\alpha}T_h^{(3-\theta)}
e^{-{1\over16\pi T_h^2}}.
\label{qbhn2} 
\end{equation}

We find that the ``dark-energy'' density  $\rho_h(t)$ (\ref{crho1}), cosmological constant (\ref{cosconst1})
and the number-density of QBHs increase, when the Universe is in radiation domination ($t>t_h, T<T_h$), 
and asymptotically approaches the constants (\ref{cosm3}) and (\ref{qbhn2}), when the Universe is in matter 
domination ($ t\gg t_h, T\ll T_h$). The numerical values of these constants (\ref{cosm3}) and (\ref{qbhn2}) 
depend on three parameters $T_h, t_h$ and $\theta$, in particular, crucially and sensitively  depends 
on the reheating temperature $T_h$ via the exponential function $\exp -1/(16\pi T_h^2)$. 

With the reheating time $t_h\simeq 113$, the present time $t\sim 10^{61}$, the index $\alpha=1/2$ 
in the radiation dominant phase, and $\theta=203/45$ for the particle content of the Standard Model, we obtain 
\begin{equation}
\rho_h \simeq  6.7 \cdot 10^{-120} m_p^4; \quad \Lambda  \simeq  1.68 \cdot 10^{-118} m_p^2,
\label{results} 
\end{equation}
by setting reheating temperature $T_h\simeq 8.45\cdot 10^{-3}$. 
This is self-consistent within this theoretical model and agrees with present observations
(\ref{lambdaobs},\ref{edobs}). The corresponding 
residual QBHs number-density is 
\begin{equation}
n_h  \simeq  2.67 \cdot 10^{-22}/{\rm cm}^3,
\label{qbhn3} 
\end{equation}
which is rather small so that the annihilation probability of residual QBHs can be neglected at the present time. The
total number ${\mathcal N}$ of residual QBHs is an approximate constant in the present evolution of Universe.
 
\section {\it Equation of state}

As Universe adiabatically expands, the internal energy $E$ of particles 
decreases, following the energy-conservation,
\begin{equation}
p\delta V+\delta E =0, 
\label{vvp} 
\end{equation}
where $p$ is the pressure, $\delta V>0$ is the variation of Universe volume and $\delta E<0$ 
the variation of internal energy of particles.
The Equation of State is 
\begin{equation}
p=(\gamma -1)\epsilon,
\label{esp} 
\end{equation}
where $\epsilon$ is the energy-density in the co-moving frame and thermal index $\gamma>1$. 
Being different from particles moving upon the manifold of four-dimension space-time, residual QBHs
are geometrical holes embed in the manifold. 
When the manifold is stretched in the expansion of Universe, these geometric holes are stretched and their sizes
become larger, as we will see, their energies increase, rather than decrease.
This implies that the ``Equation of State'' of residual QBHs must be 
completely different from the equation of state (\ref{esp}) of particles. 

The radius and mass of residual QBHs are $R=2M$ and $M=1/(8\pi T)$ ($T\simeq m_p$). 
Their 3-dimensional surface and volume are ${\mathcal A}=4\pi R^2$ and ${\mathcal V}=4\pi R^3/3$. As  
the manifold is stretched in the expansion of Universe, the radius $R$, 
the surface ${\mathcal A}$ and volume ${\mathcal V}$ of QBHs are stretched to be larger. The volume 
variation of a residual QBH is, 
\begin{equation}
\delta {\mathcal V}={\mathcal A}\delta R=2{\mathcal A}\delta M >0,
\label{vvariation} 
\end{equation}
where $\delta R$ and $\delta M$ are the variations of residual QBHs' radius and mass. 
The thermal energy (kinetic energy) 
$\epsilon_k$ of residual QBHs can be neglected, compared with their masses at the Planck scale.
Eq. (\ref{vvariation}) shows that $\delta R >0$ and $\delta M >0$ for $\delta {\mathcal V}>0$, 
indicating residual QBHs increase their size and mass-energy in the expansion of Universe.  
The energy-conservation for a residual QBH is,
\begin{equation}
p\delta {\mathcal V}+\delta M =0,
\label{vvq}
\end{equation}
in the adiabatic expansion of Universe. 
In order to obtain `` Equation of State '' of residual QBHs, we assume that 
\begin{itemize}
\item residual QBHs are uniformly distributed in the entire manifold of Universe;
\item the manifold is uniformly stretched in the expansion of Universe.
\end{itemize}
Therefore, in the expansion of universe, the energy variation of ${\mathcal N}$ residual QBHs 
is $\delta ({\mathcal N}M)={\mathcal N}\delta M$ and the volume variation of residual QBHs is 
$\delta({\mathcal N}{\mathcal V})={\mathcal N}{\delta \mathcal V}$. 
The energy-density of residual QBHs is then given by,
\begin{equation}
\epsilon=\frac {\delta ({\mathcal N}M)}{\delta({\mathcal N}{\mathcal V})}
=\frac {\delta M}{\delta \mathcal V}=\frac {1}{2\mathcal A}> 0,
\label{eqbh} 
\end{equation} 
where the thermal energy (kinetic energy) of residual QBHs' is neglected.
From Eqs. (\ref{vvq},\ref{eqbh}), we obtain the `` Equation of State '' of residual QBHs,
\begin{equation}
p=-\epsilon.
\label{eos} 
\end{equation}
Taking into account the small correction due to the thermal energy (kinetic energy) of residual QBHs, 
we can write the Equation of State (\ref{eos}) 
as $p=-(1-c)\epsilon$, where $c$ is the radio of residual QBHs' thermal energy and mass-energy, and $c \ll 1$.
The small parameter $c$ must be positive, since the thermal energy (kinetic energy) 
$\epsilon_k$ of residual QBHs decreases, i.e., $\delta \epsilon_k < 0$, as Universe volume 
expands $\delta {\mathcal V}>0$. 
 
\section {\it Summary and some remarks}

In an empty space and at zero temperature, classical particles' energy-momentum tensor (\ref{cgt1})
$T_{ab}=0$, and has no contribution to the vacuum Einstein equation (\ref{e5}).
We have argued in Sec. (\ref{vacuum0}) that quantum fields' vacuum energy-momentum (\ref{cgt}) 
$\langle 0|\hat T^V_{ab}|0\rangle=0$, and
has not contribution
to the source term (r.h.s.) of Einstein equation (\ref{e2}). This is in accordance with  
the positive-energy theorem \cite{yao}, which actually requires zero mass-energy 
in the Minkowski spacetime, and the ground state is the flat Minkowski spacetime.
This implies that the vacuum energy of quantum fields is not gravitating and thus the cosmological 
term is very unlikely related to the vacuum energy of quantum fields. 

In this article, we study the possibility that the cosmological term is given by the energy-density of 
QBHs, which is related to the finite action of classical gravitational instantons. The nucleation of QBHs is 
attributed to the unstable quantum fluctuations about the classical gravitational instantons, i.e., 
the decay of flat space at finite temperature. Based on such cosmological term, 
we study the evolution of the early Universe and present value of the cosmological constant. These are 
still very preliminary results, and there are many questions for further studies in future.



\begin{thebibliography}{99}

\bibitem{linde} 
A.H.~Guth, Phys.~Rev.~D.~23 (1981) 347;\\
A.D.~ Linde, Phys.~Lett.~B108 (1982) 389, {\it ibid} B129 (1983) 177;\\
D.~La and P.J.~Steinhardt, Phys.~Rev.~Lett.~62 (1989) 376. 

\bibitem{wein}
Ya.~B.~Zeldovich, JETP Lett., 6, 883, 1967, Sov.~Phys.~- Uspekhi 11 
(1968) 381; \\
S.~Weiberger, Rev.~Mod.~Phys.~61 (1989) 1 and Phys.~Rev.~ D61 (2000)
103505;\\
A.~Sahni and A.~Starobinsky, Int.~J.~Mod.~Phys.~D Vol.~9 No.~4, 2000;\\
V.G.Gurzadyan, S.-S.~Xue, Mod.Phys.Lett. A18 (2003) 561-568; \\
S.M.~Carroll, Contribution to Measuring and Modeling the Universe, 
Carnegie Observatories Astrophysics Series Vol. 2, ed. W. L. Freedman, 
(astro-ph/0310342), and references therein.

\bibitem{q}
P.J.E.~Peebles and B.~Ratra, Astrophys.~J 325 (1988) 17 and 
Phys.~Rev.~D 37 (1988) 3406; \\
C.~Wetterich, Nucl.~Phys.~B302 (1988) 668;\\
R.~R.~Cardwell, R.~Dave, and P.~J.~Seinhardt, Phys.~Rev.~Lett.~ 80, (1998) 1582;\\
I.~Zlatev, L.~Wang, and P.~J.~Seinhardt, Phys.~Rev.~Lett.~ 80 (1999) 896,
1998 and Phys.~Rev.~D59 (1999) 123504;\\
R.~Riotto and D.H.~Lyth, Phys.~Rept.~314 (1999)1;\\
J.~Ellis, Lectures at the Australian National University Summer School on the New Cosmology, January 2003,
(astro-ph/0305038);\\
W.H.~Kinney, Lectures given at the NATO Advanced Study Institute on Techniques and Concepts of High Energy Physics, St. Croix, USVI (2002), (astro-ph/0301448), references therein.

\bibitem{s} 
A.D.~Linde, Phys.Scripta T117 (2005) 40-48, (hep-th/0402051), references therein.

\bibitem{casimir}
H.B.G.~Casimir, Proc.~Kon.~Ned.~Akad.~Wetenschap., ser.~B, vol.~51, p.~793, 
1948;\\
see also M.~Fierz, Helv.~Phys.~Acta, vol.~33, p.~855, 1960.

\bibitem{hawking}
S.~W.~Hawking, Nature 238 (1974) 30;\\ 
S.~W.~Hawking, Commun.~Math.~Phys.~43, 199 (1975);\\ 
G.~W.~Gibbons and S.~W.~Hawking, Phys.~Rev.~D 15 2752 (1977).

\bibitem{xue2}
S.-S.~Xue, Mod.~Phys.~Lett.~A 18 (2003) 1325; Gen.~Rel.~Gravit.~37 (2005) 857; {\it ibid} 38 (2006) 1135.

\bibitem{gross}
D.J.~Gross, M.J.~Perry and L.G.~Yaffe, Phys.~Rev.~D.~25 (1982) 330.

\bibitem{gross27}
S.W.~Hawking, in {\it Recent Developments in Gravitation}, edited by M. Levy (Plenum, New York, 1979).

\bibitem{yao}
R.~Shoen and S.T.~Yau, Phys.~Rev.~Lett.~42 (1979) 547, 
Commun.~Math.~Phys.~65 (1979) 45, {\it ibid} 79 (1981) 231. 

\bibitem{gibbons0}
G.W.~Gibbons and M.J. Perry, Proc.~R.~Soc.~ London {\bf A358} (1978) 467.

\bibitem{gibbons1}
G.W.~Gibbons and S.W.~Hawking, Phys.~Rev.~D {\bf 15} (1977) 2752.

\bibitem{gibbons2}
G.W.~Gibbons and S.W.~Hawking, Commun.~Math.~Phys. {\bf 66} (1979) 291.

\bibitem{lapedes80}
A.S. Lapedes, Phys.~ Rev.~ D22 (1980) 1837.

\bibitem{belavin1975}
A.~A.~Belavin {\it et al.}, Phys.~Lett., B59 (1975) 285.

\bibitem{thooft1976} 
G.~t'Hooft, Phys.~Rev.~Lett.~37 (1976) 8, Phys.~Rev.~D {\bf 12} (1976) 3432.

\bibitem{page}
An Argument for the existence of such modes was previously given by
D.N. Page (unpublished).

\bibitem{unstable}
G.~Preparata, S.~Rovelli, and S.-S.~Xue, Phys.~Lett., B427 (1998) 254 
and Gen.~Rel.~Gravit.~ Vol 32, No. 9, (2000).

\bibitem{affleck}
I.~Affleck, Phys.~Rev.~Lett.~46 (1981) 388.

\bibitem{rate}
M.J.~Perry, Nucl.~Phys.~B143 (1978) 114;\\
S.M.~Christensen and M.J.~Duff, {\it ibid} B154 (1979) 301.

\bibitem{weinbergbookfield}
S. Weinberg, ''The Quantum Theory of Fields'', Vol. II, Cambridge University Press, (1996).

\bibitem{accretion}
B.~Carter, G.W.~Gibbons, D.N.C.~Lin and M.J.~Perry, Astron.~Astrophys. {\bf 52} 
(1976) 427.

\end{thebibliography}
\end{document}